\documentclass[seceqn,dvips]{arxstspdf}
\usepackage{graphics}
\usepackage{flushend}
\usepackage{stfloats}

\volume{24}
\issue{4}
\pubyear{2009}
\firstpage{451}
\lastpage{471}
\doi{10.1214/09-STS307}

\makeatletter
\newtheorem{lemma}{Lemma}

\makeatother

\begin{document}
\begin{frontmatter}

\title{Population Structure and Cryptic Relatedness in Genetic\\ Association Studies}
\runtitle{Population Structure}

\begin{aug}
\author[a]{\fnms{William} \snm{Astle}\ead[label=e1]{wja@ic.ac.uk}}\and
\author[b]{\fnms{David J.} \snm{Balding}\corref{}\ead[label=e2]{d.balding@ucl.ac.uk}\thanksref{tz}}
\thankstext{tz}{Current Address: Institute of Genetics, University
College London, 5 Gower Place, London, WC1E 6BT, UK.}
\runauthor{W. Astle and D. J. Balding}

\affiliation{Imperial College London}

\address[a]{William Astle is Research Associate, Centre for Biostatistics,
Department of Epidemiology and Public Health, St. Mary's Hospital Campus,
Imperial College London, Norfolk Place, London, W2 1PG, UK
\printead{e1}.}
\address[b]{David J. Balding is Professor of Statistical Genetics, Centre for Biostatistics, Department
of Epidemiology and Public Health, St. Mary's Hospital Campus,
Imperial College London, Norfolk Place, London, W2 1PG, UK \printead{e2}.}

\end{aug}

%
\begin{abstract}
We review the problem of confounding in genetic association studies,
which arises principally because of
population structure and cryptic relatedness. Many treatments of the
problem consider only a simple ``island''
model of population structure. We take a broader approach, which views
population structure and cryptic
relatedness as different aspects of a single confounder: the unobserved
pedigree defining the (often distant)
relationships among the study subjects. Kinship is therefore a central
concept, and we review methods of
defining and estimating kinship coefficients, both pedigree-based and
marker-based. In this unified framework
we review solutions to the problem of population structure, including
family-based study designs,
genomic control, structured association, regression control, principal
components adjustment and
linear mixed models. The last solution makes the most explicit use of
the kinships among the study subjects,
and has an established role in the analysis of animal and plant
breeding studies. Recent computational
developments mean that analyses of human genetic association data are
beginning to benefit from its powerful
tests for association, which protect against population structure and
cryptic kinship, as well as
intermediate levels of confounding by the pedigree.
\end{abstract}

%
\begin{keyword}
\kwd{Cryptic relatedness}
\kwd{genomic control}
\kwd{kinship}
\kwd{mixed model}
\kwd{complex disease genetics}
\kwd{ascertainment}.
\end{keyword}

\end{frontmatter}

\section{Confounding in Genetic Epidemiology}

%
\begin{figure*}[b]

\includegraphics{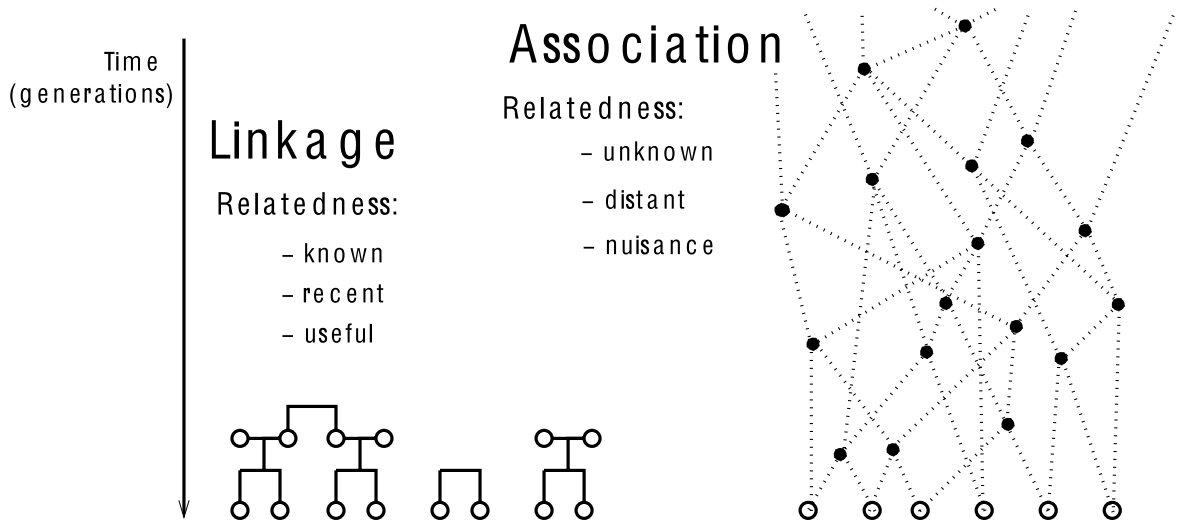}

\caption{Schematic illustration of differences between
linkage studies, which track transmissions in known pedigrees, and
population association studies which assume ``unrelated'' individuals.
Open circles denote study subjects for whom phenotype data are
available and solid lines denote observed parent-child relationships.
Dotted lines indicate unobserved lines of descent, which may extend
over many generations, and filled circles indicate the common ancestors
at which these lineages first diverge. Unobserved ancestral lineages
also connect the founders of a linkage study, but these have little
impact on inferences and are ignored, whereas they form the basis of
the rationale for an association analysis and constitute an important
potential confounder.}\label{linkass}
\end{figure*}

\subsection{Association and Linkage}\label{chasstpopstruc}

Genetic \textit{association studies} (Clayton, \citeyear{clayhbook}) are designed to
identify genetic loci at which the allelic state
is correlated 
with a phenotype of interest. The associations of interest
are causal, arising at loci whose different alleles have different
effects on phenotype.
Even if a causal locus is not genotyped in the
study, it may be possible to identify an association indirectly through
a genotyped locus that is nearby on the genome. In this review we are
concerned with the task of guarding against spurious associations,
those which do not arise at or near a causal locus. We first introduce
background material describing linkage and association studies,
population structure and linkage disequilibrium, the problem of
confounding by population structure and cryptic relatedness. In Section~\ref{secrelation} we discuss definitions and estimators of the kinship coefficients
that are central to our review of methods of correcting for confounding
by population structure and cryptic relatedness, which is presented in
Section~\ref{sec3}. Finally, in Section~\ref{sim} we present the results of a small
simulation study illustrating the merits of the most important methods
introduced in Section~\ref{sec3}.

Although association designs are used to study other species, we will
mainly take a human-genetics viewpoint. For example, we will focus on
binary phenotypes, such as disease case/control or drug
responder/nonresponder, which remain the most commonly studied type of
outcome in humans, although quantitative (continuous), categorical and
time-to-event traits are increasingly important. The subjects of an
association study are sometimes sampled from a population without
regard to phenotype, as in prospective cohort designs. However,
retrospective ascertainment of individuals on the basis of phenotype,
as in case-control study designs, is more common in human genetics, and
we will focus on such designs here.

\textit{Linkage studies} (Thompson, \citeyear{thomhbook}) form the other major class of
study designs in genetic epidemiology. These seek loci at which there
is correlation between the phenotype of interest and the pattern of
transmission of DNA sequence over generations in a known pedigree. In
contrast, association studies are used to search for loci at which
there is a significant association between the phenotypes and genotypes
of unrelated individuals. These associations arise because of
correlations in transmissions of phenotypes and genotypes over many
generations, but association analyses do not model these transmissions
directly, whereas linkage analyses do. The relatedness of study
subjects is therefore
central to a linkage study, whereas the
relatedness of association study subjects is typically unknown and
assumed to be distant; any close relatedness is a nuisance (Figure~\ref{linkass}).

In the last decade, association studies have become increasingly
prominent in human genetics, while, although they remain important, the
role of linkage studies has declined. Linkage studies can provide
strong and robust evidence for genetic causation, but are limited by
the difficulty of ascertaining enough suitable families, and by
insufficient recombinations within these families to refine the
location of a causal variant. When only a few hundred of genetic
markers were available, lack of within-family recombinations was not a
limitation. Now, cost-effective technology for genotyping $\sim$10$^6$
\textit{single nucleotide polymorphism} (SNP) markers distributed across
the genome has made possible genome-wide association studies (GWAS)
which investigate most of the common genetic variation in a population,
and obtain orders of magnitude finer resolution than a comparable
linkage study (Morris and Cardon, \citeyear{cardhbook};
Altshuler, Daly and Lander, \citeyear{DavidAltshuler11072008}). GWAS are
preferred for detecting common causal variants (say, population
fraction $>$ 0.05), which typically have only a weak effect on
phenotype, whereas linkage studies remain superior for the detection of
rare variants of large effect (because these effects are more strongly
concentrated within particular families).

%
\begin{figure}[b]

\includegraphics{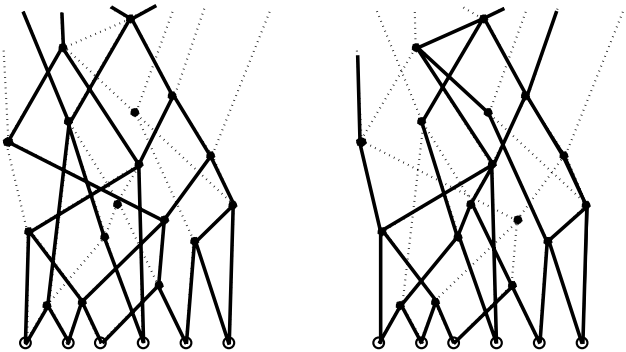}

\caption{Schematic illustration of the confounding role
of pedigree on ancestral lineages at individual loci. Two possible
single-locus lineages are shown (solid lines), each embedded in the
pedigree of Figure~\protect\ref{linkass} (right). Moving upwards from the study
subjects (open circles), when two lineages meet at a common ancestor
(filled circle), they either coalesce into a single lineage, or else
they pass through different alleles of the common ancestor and do not
coalesce. Dotted lines show pedigree relationships that do not
contribute to the ancestry of the study subjects at this locus.
Although lineages are random, they are constrained by the pedigree,
features of which are therefore reflected in lineages across the genome.}\label{genped}
\end{figure}

Because genes are essentially immutable during an individual's
lifetime, and because of the independence of allelic transmissions at
unlinked loci (Mendel's Second Law), linkage studies are virtually
immune to confounding. Association studies are, however, susceptible to
genetic confounding, which is usually thought of as coming in two
forms: \textit{population structure} and \textit{cryptic
relatedness}. These
are in fact two ends of a spectrum of the same confounder: the
unobserved pedigree specifying the (possibly distant) relationships
among the study subjects (Figure~\ref{linkass}, right). Association
studies are also susceptible to confounding if genotyping error rates
vary with phenotype (Clayton et~al., \citeyear{PubMed16228001}). This can resemble a form
of population structure and is not discussed further here.

We can briefly encapsulate the genetic confounding problem as follows.
Association studies seek genomic loci at which differences in the
genotype distributions between cases and controls indicate that their
ancestries are systematically different \textit{at that locus}. However,
pedigree structure can generate a tendency for systematic ancestry
differences between cases and controls at all loci not subject to
strong selection. Figure~\ref{genped} illustrates two possible
ancestral lineages of the study subject alleles at a locus. Lineages
are correlated because they are constrained to follow the underlying
pedigree. For example, if the pedigree shows clustering of individuals
into subpopulations, then ancestral lineages at neutral loci will tend
to reflect this. The goal of correction for population structure is to
allow for the confounding pedigree effects when assessing differences
in ancestry between cases and controls at individual loci. In the
following sections we seek to expand on this brief characterization.

\subsection{Population Structure}\label{popstruc}
Informally, a population has structure when there are large-scale
systematic differences in ancestry, for example, varying levels of
immigrant ancestry, or groups of individuals with more recent shared
ancestors than one would expect in a \textit{panmictic} (random-mating)
population. Shared ancestry corresponds to relatedness, or \textit{
kinship}, and so population structure can be defined in terms of
patterns of kinship among groups of individuals. Population structure
is often closely aligned with geography, and in the absence of genetic
information, stratification by geographic region may be employed to try
to identify homogeneous subpopulations. However, this approach does not
account for recent migration or for nongeographic patterns of kinship
based on social or religious groups.

The simplest model of population structure assumes a partition of the
population into ``islands'' (subpopulations). Mating occurs
preferentially between pairs of individuals from the same island, so
that the island allele fractions tend to diverge to an extent that
depends on the inter-island migration rates. An enhancement of the
island model to incorporate admixture allows individual-specific
proportions of ancestry arising from actual or hypothetical ancestral islands.

Below we will focus on island models of population structure, because
these are simple and parsimonious models that facilitate discussion of
the main ideas. Moreover, several popular statistical methods for
detecting population structure and correcting association analysis for
its effects have been based entirely on such models. However, human
population genetic and demographic studies suggest that island models
typically do not provide a good fit for human genetic data.
Colonization often occurs in waves and is influenced by geographic and
cultural factors. Such processes are expected to lead to clinal
patterns of genetic variation rather than a partition into
subpopulations (Handley et~al., \citeyear{PubMed17655965}). Modern humans are known to have
evolved in Africa with the first wave of human migration from Africa
estimated to have been approximately 60,000 years ago. Reflecting this
history, current human genetic diversity decreases roughly linearly
with distance from East Africa (Liu et~al., \citeyear{PubMed16826514}). Within Europe,
Lao et al. (\citeyear{PubMed18691889}) found that the first two principal components of
genome-wide genetic
variation accurately reflect latitude and
longitude: there is population structure at a Europe-wide level, but no
natural classification of Europeans into a small number of
subpopulations. Similarly, there does not appear to be a simple
admixture model based on hypothetical ancestral subpopulations that can
adequately capture European genetic variation, although a model based
on varying levels of admixture from hypothetical ``North Europe'' and
``South Europe'' subpopulations could at least capture the latitude
effect. The admixture model may be appropriate when the current
population results from some intermixing following large-scale
migrations over large distances, such as in Brazil or the Caribbean.

Because the term ``population stratification'' can imply an underlying
island model, we avoid this term and adhere to ``population
structure,'' which allows for more complex underlying demographic models.

\subsection{Linkage Disequilibrium}

\begin{figure}

\includegraphics{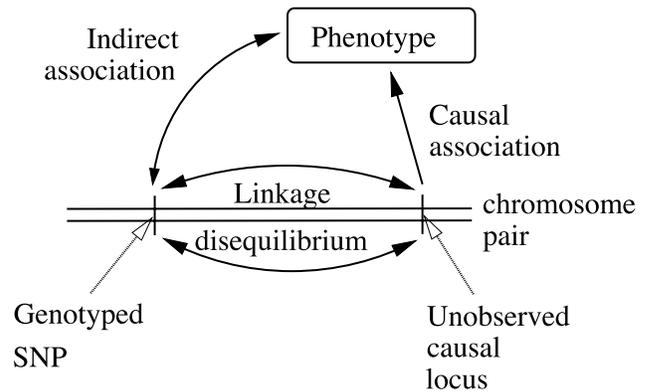}

\caption{Illustration of the role of linkage disequilibrium
in generating phenotypic association with a noncausal genotyped marker
due to a tightly-linked ungenotyped causal locus.}\label{ass}
\end{figure}

In a large, panmictic population, and in the absence of selection,
pairs of genetic loci that are not \textit{tightly linked} (close
together on a chromosome) are unassociated at the population level
(McVean, \citeyear{mcveanhb}). Such \textit{linkage equilibrium} arises because
recombination events ensure the independent assortment of alleles when
they are transmitted across generations (a process sometimes called
Mendelian Randomization). Conversely, because recombination is rare
($\sim$1 recombination per chromosome per generation),\break tightly linked
loci are generally correlated, or in \textit{linkage disequilibrium}
(LD) in the population. This is because many individuals can inherit a
linked allele pair from a remote common ancestor without an intervening
recombination. Association mapping relies on LD because, even for a
GWAS, only a small proportion of genetic variants are directly
measured. Signals from ungenotyped causal variants can only be detected
through phenotype association with a genotyped marker that is in
sufficiently strong LD with the causal variant (Figure~\ref{ass}). LD
is a double-edged sword: the stronger the LD around a causal variant,
the easier it is to detect, because the greater the probability it is
in high LD with at least one genotyped marker (Pritchard and Przeworski, \citeyear{PubMed11410837}).
However, in a region of high LD it is hard to fine-map a causal variant
because there will be multiple highly-correlated markers each showing a
similar strength of association with the phenotype.

%

\subsection{Spurious Associations due to\break Population Structure}
Unfortunately, population structure can cause LD between unlinked loci
and consequently generate spurious marker-phenotype associations. For
example, in the island model of population structure, if the proportion
of cases among the sampled individuals varies across subpopulations,
then alleles that vary in frequency across subpopulations will often
show association with phenotype. One or more such alleles may in fact
be involved in phenotype determination, but standard association
statistics may not distinguish them from the many genome-wide alleles
with frequencies that just happen to vary across subpopulations because
of differential genetic drift or natural selection. To express this
another way, many alleles across the genome are likely to be somewhat
informative about an individual's subpopulation of origin, and hence be
predictive of any phenotype that varies across subpopulations. For
example, in a large sample drawn from the population of Great Britain,
many genetic variants are likely to show association with the phenotype
``speaks Welsh.'' These will be alleles that are relatively common in
Wales, which has a different population history from England (Weale et~al., \citeyear{PubMed12082121}), and do not ``cause'' speaking Welsh.

Under an island model, one could potentially solve the problem of
spurious associations by matching for ancestry, for example, by
choosing for each case a control from the same subpopulation. However,
as noted above, an island model is unlikely to describe the ancestry of
a human population adequately. We each have a distinct pattern of
ancestry, to a large extent unknown beyond a few generations, making
precise matching impractical while crude matching may be insufficient.
The spouse of a case, or another relative by marriage, can provide a
genetically unrelated control approximately matched for ancestry, but
there are obvious limitations to this approach.

There are at least three reasons why, in an unmatched study, the
phenotypes of study subjects might vary systematically with ancestry
(e.g.,\ with subpopulation in an island model). The most
straightforward reason is that the disease prevalence varies across
subpopulations in accordance with the frequencies of causal alleles,
and the differing sample case:control ratios across subpopulations
reflect the differing subpopulation prevalences. Alternatively,
subpopulation prevalences may vary because of differing environmental
risks. Third, \textit{ascertainment bias} can make an important
contribution to associations between ancestry and phenotype.
Ascertainment bias can arise if there are differences in the sampling
strategies between cases and controls that are correlated with
ancestry. In the island model, this means that the sample case:control
ratios across subpopulations do not reflect the subpopulation
prevalences. This may happen, for example, because cases, but not
controls, are sampled from clinics that over-represent particular groups.

\subsection{Extent of the Problem}\label{extent}
The vulnerability of association studies to confounding by population
structure has been recognized for many years. In a famous example,
Knowler et~al. (\citeyear{PubMed3177389}) found a significant association between an
immunoglobulin haplotype and type II diabetes. The study subjects were
native North Americans with some European ancestry and the association
disappeared after stratification by ancestry. Many commentators fail to
note that Knowler et al. understood the problem and performed
an appropriate analysis, so that no false association was reported:
they merely noted the potential for confounding in an unstratified analysis.

Marchini et~al. (\citeyear{PubMed15052271}) concluded from a simulation study that, even in
populations with relatively modest levels of structure (such as Europe
or East Asia), when the sample is large enough to provide the required
power, the most significant SNPs can have their \textit{p}-values
reduced by a factor of three because of population structure, thus
exaggerating the significance of the association. Freedman et~al. (\citeyear{PubMed15052270}) examined a study into prostate cancer in (admixed)
African Americans and estimated a similar reduction in the smallest
\textit{p}-values. Another study of European-Americans found a SNP in
the lactase gene significantly associated with variation in height
(Campbell et~al., \citeyear{PubMed16041375}). When the subjects were stratified according to
North/West or South/East European ancestry, the association
disappeared. Since we expect connections among lactase tolerance, diet
and height, the association could be genuine and involve different
diets, but the confounding with population structure makes this
difficult to establish. Helgason et~al. (\citeyear{PubMed15608637}) used pedigree and marker
data from the Icelandic population, and found evidence of population
structure in rural areas, which would result on average in a 50\%
increase in the magnitude of a $\chi^2_1$ association statistic.

Following Pritchard and Rosenberg (\citeyear{PubMed10364535})
and Gorroochurn et~al. (\citeyear{PubMed15604563}), Rosenberg and Nordborg (\citeyear{PubMed16582435}) considered a general model for populations with
continuous and discrete structure and presented necessary and
sufficient conditions for spurious association to occur at a given
locus. They defined a parameter measuring the severity of confounding
under general ascertainment schemes, and showed that, broadly speaking,
the case of two discrete subpopulations is worse than the cases of
either more subpopulations or an admixed population. As the number of
subpopulations becomes larger, the problem of spurious association
tends to diminish because the law of large numbers smoothes out
correlation between disease risk and allele frequencies across
subpopulations (Wang, Localio and Rebbeck, \citeyear{PubMed15185399}).

In recent years results have been published from hundreds of GWAS into
complex genetic traits\break ({NHGRI GWAS Catalog, \citeyear{genomewidelist}). McCarthy et~al. (\citeyear{PubMed18398418})
described the current consensus. The impact of population structure on
association studies should be modest ``as long as cases and controls
are well matched for broad ethnic background, and measures are taken to
identify and exclude individuals whose GWAS data reveal substantial
differences in genetic background.'' This is consistent with a report
from a study of type II diabetes in UK Caucasians which estimated that
population structure was responsible for only $\sim$4\% inflation in
$\chi^2_1$ association statistics (Clayton et~al., \citeyear{PubMed16228001}).
{The Wellcome Trust Case Control Consortium (\citeyear{PubMed17554300}) study of seven common diseases using a UK population
sample found fewer than 20 loci exhibiting strong geographic
variation. The genome-wide distribution of test statistics suggested
that any confounding effect was modest and no adjustment for population
structure was made for the majority of their analyses.

In conclusion, the magnitude of the effect of structure depends on the
population sampled and the sampling scheme, and well-designed studies
should usually suffer only a small impact. However, most of the
associated variants so far identified by GWAS have been of small effect
size ({NHGRI GWAS Catalog, \citeyear{genomewidelist}), and as study sizes increase in order to
detect smaller effects, even modest structure could substantially
increase the risk of false positive associations.

\subsection{Cryptic Relatedness}
Cryptic relatedness refers to the presence of close relatives in a
sample of ostensibly unrelated individuals. Whereas population
structure generally describes remote common ancestry of large groups of
individuals, cryptic relatedness refers to recent common ancestry among
smaller groups (often just pairs) of individuals. Like population
structure, cryptic relatedness often arises in unmatched association
studies and can have a confounding effect on inferences.
Indeed, Devlin and Roeder (\citeyear{PubMed11315092}) argued that cryptic relatedness could pose a more
serious confounding problem than population structure. A subsequent
theoretical investigation of plausible demographic and sampling
scenarios (Voight and\break Pritchard, \citeyear{PubMed16151517}) showed that the effect of cryptic
relatedness in well-designed studies of outbred populations should be
negligible, but it can be noticeable for small and isolated
populations. Using pedigree and empirical genotype data from the
Hutterite population, these authors found that cryptic relatedness
reduces an association \textit{p}-value of $10^{-3}$ by a factor of
approximately 4, and that the smaller the \textit{p}-value the greater
is the relative effect.

\section{Genetic Relationships}\label{secrelation}

\subsection{Kinship Coefficients Based on\break Known Pedigrees}

The relatedness between two diploid individuals can be defined in terms
of the probabilities that each subset of their four alleles at an
arbitrary locus is \textit{identical by descent} (IBD), which means
that they descended from a common ancestral allele without an
intermediate mutation. The probability that the two homologous alleles
within an individual $i$ are IBD is known as its \textit{inbreeding
coefficient}, $f_i$. When no genotype data are available, IBD
probabilities can be evaluated from the distribution of path lengths
when tracing allelic lineages back to common ancestors (Figure~\ref{genped}), convolved with a mutation model (Mal{\'{e}}cot, \citeyear{malecot}). More
commonly, IBD is equated with ``recent'' common ancestry, where
``recent'' may be defined in terms of a specified, observed pedigree,
whose founders are assumed to be completely unrelated. In theoretical
models, ``recent'' may be defined, for example, in terms of a specified
number of generations, or since the last migration event affecting a
lineage. Linkage analysis conditions on the available pedigree, and in
this case the definition of IBD in terms of shared ancestry within that
pedigree, and the assumption of unrelated founders, cause no
difficulty. However, the strong dependence on the observed pedigree, or
other definition of ``recent'' shared ancestry, is clearly
unsatisfactory for a more general definition of relatedness.

%
\begin{figure}

\includegraphics{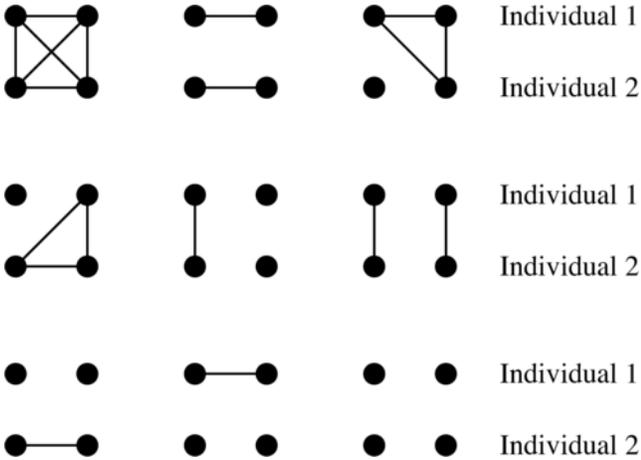}

\caption{Schematic illustration of the nine relatedness
classes for two individuals, whose four alleles are indicated by filled
circles, that are specified by the eight Jacquard identity-by-descent
(IBD) coefficients. Within-individual allele pairs are regarded as
unordered, and solid lines link alleles that are IBD.}\label{jac}
\end{figure}

A full description of the relatedness between two diploid individuals
requires $15$ IBD probabilities, one for each nonempty subset of four
alleles, but if we regard the pair of alleles within each individual as
unordered, then just eight identity coefficients (Jacquard, \citeyear{jacq}) are
required (Figure~\ref{jac}). An assumption of no within-individual IBD
(no inbreeding) allows these eight coefficients to be collapsed into
two (Cotterman, \citeyear{cott40}), specifying probabilities for the two individuals to
share exactly one and two alleles IBD. Both these coefficients are
required for models involving dominance, but for additive genetic
models they can be reduced to a single \textit{kinship coefficient},
$K_{ij}$, which is the probability that two alleles, one drawn at
random from each of $i$ and $j$, are IBD. Similarly, $K_{ii}$ is the
probability that two alleles, sampled with replacement from $i$, are
IBD. Thus, $K_{ii}=(1+f_i)/2$, and, in particular, the kinship of an
outbred individual with itself is $1/2$.

The kinship matrix $K$ of a set of individuals in a pedigree can be
computed by a recursive algorithm that neglects within-pedigree
mutation (Thompson, \citeyear{thom}). $K$ is positive semi-definite if the submatrix
of assumed founder kinships is positive semi-definite (which is
satisfied if, as is typical, founders are assumed unrelated).

\subsection{Kinship Coefficients Based on Marker Data}\label{kinmarker}

The advent of GWAS data means that genome-average relatedness can now
be estimated accurately. It can be preferable to use these estimates in
association analyses even if (unusually) pedigree-based estimates are
available. There is a subtle difference between expectations computed
from even a full pedigree, and realized amounts of shared genomic
material. For example, if two lineages from distinct individuals meet
in a common ancestor many generations in the past, then this ancestor
will contribute (slightly) to the pedigree-based relatedness of the
individuals but may or may not have passed any genetic material to both
of them. Similarly, two pairs of siblings in an outbred pedigree may
have the same pedigree relatedness, but (slightly) different empirical
relatedness (Weir, Anderson and Hepler, \citeyear{PubMed16983373}).

Thompson (\citeyear{PubMed1052764}) proposed maximum likelihood estimates (MLEs) of
the Cotterman coefficients, while Milligan (\citeyear{PubMed12663552}) made a
detailed study of MLEs under the Jacquard model. These MLEs can be
prone to bias when the number of markers is small and can be
computationally intensive to obtain particularly from genome-wide data
sets (Ritland, \citeyear{Ritland};
Milligan, \citeyear{PubMed12663552}).

Method of moments estimators (MMEs) are typically less precise than
MLEs, but are computationally efficient and can be unbiased if the
ancestral allele fractions are known (Milligan, \citeyear{PubMed12663552}). Under
many population genetics models, if two alleles are not IBD, then they
are regarded as random draws from some mutation operator or allele pool
(Rousset, \citeyear{PubMed11986874}), which corresponds to the notion of
``unrelated.'' The kinship coefficient $K_{ij}$ is then a correlation
coefficient for variables indicating whether alleles drawn from each of
$i$ and $j$ are some given allelic type, say, $\textsf{A}$. If $x_i$
and $x_j$ count the numbers of $\textsf{A}$ alleles (0, 1 or 2) of $i$
and $j$, then
%
%
\begin{equation}\label{cov}
\operatorname{Cov}(x_i,x_j)=4p(1-p)K_{ij},
\end{equation}
where $p$ is the population fraction of $\textsf{A}$ alleles. Thus,
$K_{ij}$ can be estimated from genome-wide covariances of allele
counts. Specifically, if we write $x$ as a column vector over
individuals and let the subscript index the $L$ loci (rather than
individuals), then
%
%
\begin{equation}\label{corest}
\hat{K}=\frac{1}{L}\sum_{l=1}^L\frac{(x_l-2p_l\mathbf{1})(x_l-2p_l\mathbf{1})^T}{4p_l(1-p_l)}
\end{equation}
is an unbiased and positive semi-definite estimator for the kinship
matrix $K$. Entries in $\hat{K}$ can also be interpreted in terms of
excess allele sharing beyond that expected for unrelated individuals,
given the allele fractions. According to Ritland (\citeyear{Ritland}), who considered
similar estimators and gave a generalization to loci with more than two
alleles, (\ref{corest}) was first given in Li and Horvitz (\citeyear{PubMed13065259}) but
only for inbreeding coefficients.

In practice, we do not know the allele fractions $p_l$. The natural
estimators assume outbred and unrelated individuals, deviation from
which can exaggerate the downward bias in the $K_{ij}$ estimates that
arises from the overfitting effect of estimating the $p_l$ from the
same data. To reduce the first problem, one could iteratively
re-estimate the $p_l$ after making an initial estimate of $K$ with
\[
\hat{p}_l=\frac{\mathbf{1}^T\hat{K}^{-1}x_l}{\mathbf{1}^T\hat{K}^{-1}\mathbf{1}}.
\]
Although the correlations arising from shared ancestry are in principle
positive, because of bias arising from estimation of the $p_l$,
off-diagonal entries of (\ref{corest}) can be negative, a property that
has caused some authors to shun such estimators of $K$ (Milligan, \citeyear{PubMed12663552};
Yu et~al., \citeyear{PubMed16380716};
Zhao et~al., \citeyear{PubMed17238287}). Rousset (\citeyear{PubMed11986874}) also criticized the model underlying (\ref{cov}) in
the context of certain population genetics models, but did not propose
an alternative estimator of genetic covariance in actual populations.
For our purpose, that of modeling phenotypic correlations, genotypic
correlations seem intuitively appropriate and the interpretation of
$K_{ij}$ as a probability seems unimportant. Under the interpretation
of $\hat K_{ij}$ as excess allele sharing, negative values correspond
to individuals sharing fewer alleles than expected given the allele frequencies.

%
\begin{table*}
\caption{Identity-by state (IBS) coefficients at a single
diallelic locus, defined as the probability that alleles drawn at
random from $i$ and $j$ match, which gives $0.5$ in the case of a pair
of heterozygotes. Another definition, based on the number of alleles in
common between $i$ and $j$, gives 1 for a pair of heterozygotes}\label{ibs}
\begin{tabular*}{\textwidth}{@{\extracolsep{4in minus 4in}}l ccc ccc ccc @{}}
\hline
Genotype of
$i$&\textsf{aa}&\textsf{Aa}&\textsf{AA}&\textsf{aa}&\textsf{Aa}&\textsf{AA}&\textsf{aa}&\textsf{Aa}&\textsf{AA}\\[3pt]
Genotype of $j$&\textsf{aa}&\textsf{aa}&\textsf{aa}&\textsf{Aa}&\textsf{Aa}&\textsf{Aa}&\textsf{AA}&\textsf{AA}&\textsf{AA}\\[3pt]
IBS coefficient&$1$&$\tfrac{1}{2}$&$0$&$\tfrac{1}{2}$&$\tfrac{1}{2}$&$\tfrac{1}{2}$&$0$&$\tfrac{1}{2}$&$1$\\
\hline
\end{tabular*}
\end{table*}

Table~\ref{ibs} shows the probability that alleles chosen at random
from each of two individuals match, that is, are \textit{identical by
state} (IBS), at a genotyped diallelic locus. The genome-wide average
IBS probability can be expressed as
%
%
\begin{equation}\label{IBSmat}
\frac{1}{2L}\sum_{l=1}^L(x_l-\mathbf{1})(x_l-\mathbf{1})^T+\frac{1}{2}.
\end{equation}
If the mutation rate is low, IBS usually arises as a result of IBD, and
(\ref{IBSmat}) can be regarded as an MME of the pedigree-based kinship
coefficient in the limiting case that IBS implies IBD. This estimator
overcomes the problem with pedigree-based estimators of dependence on
the available pedigree, but it is sensitive to recurrent mutations.

Software for computing average allele sharing (IBS) is included in
popular packages for GWAS analysis such as PLINK (Purcell et~al., \citeyear{PubMed17701901}). However, because the excess allele-sharing
(genotypic correlation) estimator of kinship coefficients (\ref{corest}) incorporates weighting by allele frequency, it is typically
more precise than (\ref{IBSmat}). Sharing a rare allele suggests closer
kinship than sharing a common allele, because the rare allele is likely
to have arisen from a more recent mutation event (Slatkin, \citeyear{ageslatkin}). To
illustrate the increased precision of (\ref{corest}) over (\ref{IBSmat}), we simulated $500$ genetic data sets comprising $200$
idealized cousin pairs (no mutation, and the alleles not IBD from the
common grandparents were independent draws from an allele pool) and
$800$ unrelated individuals, all genotyped at $10{,}000$ unlinked SNPs.
After rescaling to ensure the two estimators give the same difference
between the mean kinship estimate of cousin pairs and mean kinship
estimate of unrelated pairs, the resulting standard deviations
(Table~\ref{ibsvsibd}) are about $40\%$ larger for the total allele
sharing (IBS) estimator (\ref{IBSmat}) than for the excess
allele-sharing (genetic correlation) estimator (\ref{corest}).

%
\begin{table}[b]
\caption{Estimated standard deviations of two kinship
coefficient MMEs, after linear standardization to put the estimates on\break
comparable scales}\label{ibsvsibd}
\begin{tabular*}{\columnwidth}{@{\extracolsep{\fill}}lcc@{}}
\hline
\textbf{Estimator}&\textbf{Unrelated pair}&\textbf{Cousin pair}\\
\hline
Genetic correlation (\ref{corest})&5.0&5.3\\
IBS (\ref{IBSmat})&7.3&7.2\\
\hline
\end{tabular*}
\end{table}

The marker-based estimates of kinship coefficients discussed above do
not take account of LD between markers, nor do they exploit the
information about kinship inherent from the lengths of genomic regions
shared between two individuals from a recent common ancestor (Browning, \citeyear{PubMed18430938}). Hidden Markov models provide one approach to account
for LD\break (Boehnke and Cox, \citeyear{PubMed9311748};
Epstein, Duren and\break Boehnke, \citeyear{PubMed11032786}). In outbred populations,
the IBD status along a pair of chromosomes, one taken from each of a
pair of individuals in a sibling, half sib or parent-child
relationship, is a Markov process. However, the Markovian assumption
fails for more general relationships in outbred populations. When
relationships are more distant, regions of IBD will tend to cluster.
For example, in the case of first cousins IBD regions will cluster into
larger regions that correspond to inheritance from one of the two
shared grandparents. McPeek and Sun (\citeyear{PubMed10712219}) showed how to augment the
Markov model to describe the IBD process when the chromosomes
correspond to an avuncular or first cousin pair. Despite the invalidity
of the Markov assumption, Leutenegger et~al. (\citeyear{PubMed12900793}) found that in practice
it can lead to reasonable estimates for relationships more distant than
first-degree.

\section{Correcting Association Analysis\break for Confounding}\label{sec3}

In this review, we seek to use kinship to illuminate connections among
popular methods for protecting association analyses from confounding.
Many of these methods can be formulated within standard regression
models that express the expected value of $y_i$, the phenotype of the
$i$th individual, as a function of its genotype $x_i$ at the SNP of interest:
%
%
\begin{equation}\label{pro}
g(\mathbb{E}[y_i])=\alpha+x_i\beta,
\end{equation}
where, for simplicity, we have not included covariates. Here $g$ is a
link function and $\beta$ is a scalar or column vector of genetic
effect parameters at the SNP. Often $x_i$ counts the number of copies
of a specified allele carried by $i$, or it can be a two-dimensional
row vector that implies a general genetic model.

For a case-control study, $g$ is typically the $\operatorname{logit}$
function and
$\beta$ are log odds ratios. This is a prospective model, treating
case-control status as the outcome, but inferences about $\beta$ are
typically the same as for the retrospective model, which is more
appropriate for case-control data (Prentice and Pyke, \citeyear{042862078};
Seaman and Richardson, \citeyear{113262364}).
However, in some settings ascertainment effects are not correctly
modeled prospectively, and it is necessary to consider retrospective
models of the type
%
%
\begin{equation}\label{ret}
g(\mathbb{E}[x_i])=\alpha+y_i\beta,
\end{equation}
where $g$ is typically the identity function.

\subsection{Family-Based Tests of Linkage and Association (FBTLA)}\label{FBA}

The archetypal FBTLA is the transmission disequilibrium test (TDT)
(Spielman, McGinnis and Ewens, \citeyear{PubMed8447318}) for systematic differences between the genotypes
of affected children and those expected under Mendelian randomization
of the alleles of their unaffected parents. If an allele is
directly\break
risk-enhancing, it will be over-transmitted to cases. If not directly
causal but in LD with a causal allele, it may also be over-transmitted,
but in this case it must also be linked with the causal variant, since
otherwise Mendelian randomization will eliminate the association
between causal and tested alleles. Thus, the TDT is a test for both
association and linkage. The linkage requirement means that the test is
robust to population structure, while the association requirement
allows for fine-scale localization.

Parents that are homozygous at the tested SNP are uninformative and not
used. Transmissions from heterozygote parents are assumed to be
independent, which implies a multiplicative disease model. Let
$n_{\mathsf{a}}$ and $n_\mathsf{A}$ denote respectively the number of
$\textsf{a}$ and $\textsf{A}$ alleles transmitted to children by
$\textsf{Aa}$ heterozygote parents. If there is no linkage, each
parental allele is equally likely to be transmitted, so that the null
hypothesis for the TDT is
\[
H_0{}\dvtx{} \mathbb{E}[n_\mathsf{a}]=\mathbb{E}[n_\mathsf{A}].
\]
Conditional on the number of heterozygote parents $n_\mathsf{a}+n_\mathsf{A}$, the test statistic $n_\mathsf{a}$ has a
$\operatorname{Binomial}(n_\mathsf{a}+n_\mathsf{A},\break 1/2)$ null distribution, but
McNemar's statistic
%
%
\begin{equation}\label{macstat}
\frac{(n_\mathsf{a}-n_\mathsf{A})^2}{n_\mathsf{a}+n_\mathsf{A}},
\end{equation}
which has an approximate $\chi^2_1$ null distribution\break (Agresti, \citeyear{Agresti}),
is widely used instead. The TDT can be derived from the score test of a
logistic regression model in which transmission is the outcome
variable, and the parental genotypes are predictors (Dudbridge, \citeyear{dudster}). In
Section~\ref{MCPsec} we outline a test which can exploit between-family
as well as within-family information when it is available, while
retaining protection from population structure. Tiwari et~al. (\citeyear{PubMed18382087})
survey variations of the TDT in the context of a review of methods of
correction for population structure.

The main disadvantages of the TDT and other FBTLA are the problem of
obtaining enough families for a well powered study (particularly for
adult-onset diseases) and the additional cost of genotyping: three
individuals must be genotyped to obtain the equivalent of one
case-control pair, and homozygous parents are uninformative. Given the
availability of good analysis-based solutions to the problem of
population structure (see below), the design-based solution of the
FBTLA pays too high a price for protection against spurious
associations (Cardon and Palmer, \citeyear{PubMed12598158}). However, FBTLA designs (like
other linkage designs) can also be used to investigate parent-of-origin
effects (Weinberg, \citeyear{PubMed10364536}), which is not usually possible for
population-based case-control studies.

\subsection{Genomic Control}\label{GenCon}

Genomic Control (GC) is an easy-to-apply and computationally fast
method for reducing the inflation of test statistics caused by
population structure or cryptic relatedness. It can be applied to data
of any family structure or none. GC was developed (Devlin and Roeder, \citeyear{PubMed11315092}) for the Armitage test statistic, which is
asymptotically equivalent to a score statistic under logistic
regression (Agresti, \citeyear{Agresti}) and, in the absence of confounding, has an
asymptotic $\chi_1^2$ null distribution. The Armitage test assumes an
additive disease model, but GC has also been adapted for tests of other
disease models (Zheng et~al., \citeyear{PubMed15737092};
Zheng, Freidlin and Gastwirth, \citeyear{PubMed16400614}).

%
\begin{figure*}

\includegraphics{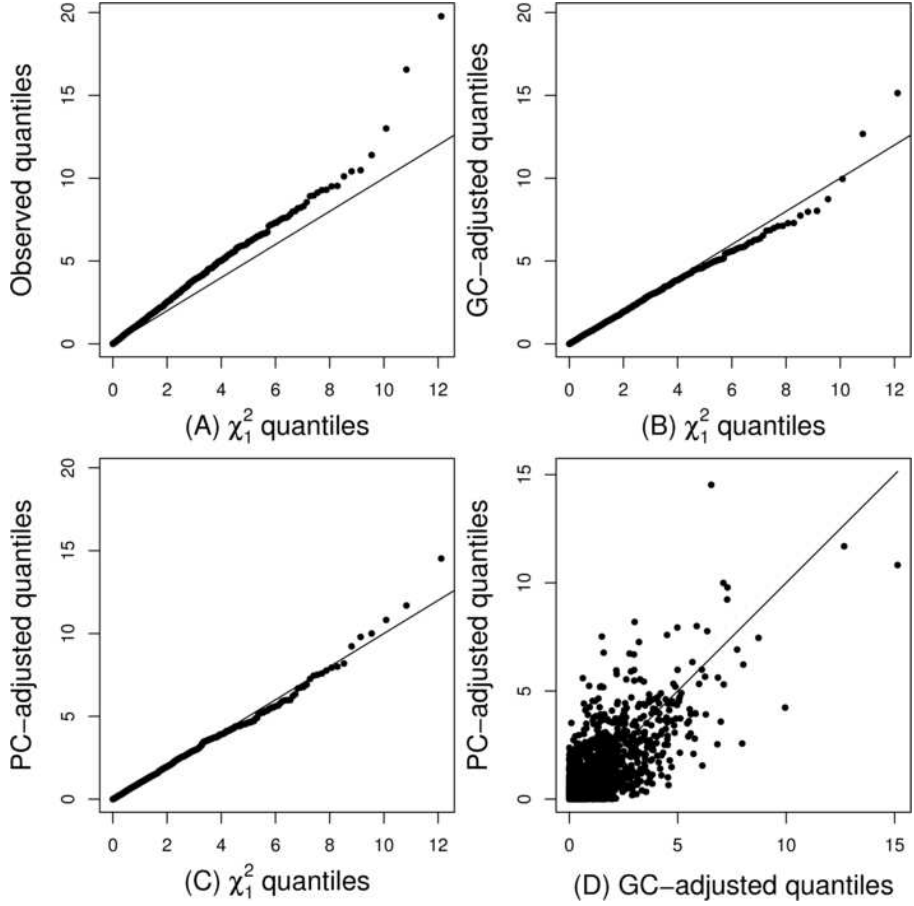}

\caption{Q--Q plots for likelihood ratio tests of
association in logistic regression (equivalent to the Armitage trend
test), at 2000 null SNPs simulated under a three-island model with
$F_{ST}=1\%$. From Island 1 there are 200 controls and 100 cases. Each
of the remaining 700~individuals is admixed, the $i$th individual
having a proportion $a_i$ of their ancestry from Island 2, the
remainder from Island 3, where the $a_i$ are independent and
$\operatorname{Uniform}(0,1)$. The $i$th admixed individual has a probability
$0.3+0.5\times a_i$ of being a case, so that case status is positively
correlated with Island 2 ancestry. \textup{(A)} expected versus observed
quantiles, unadjusted; \textup{(B)} expected versus observed after GC
median-adjustment; \textup{(C)} expected versus observed when the first two
principal components are included as covariates; \textup{(D)} GC-adjusted versus
PC-adjusted quantiles.}\label{gcpc}
\end{figure*}

Figure~\ref{gcpc}(A) illustrates the inflation of Armitage test
statistics at 2000 null SNPs simulated under an island model with
admixture and ascertainment bias. This inflation could reflect many
genome-wide true associations, but it is more plausible (and correct
for this simulation) that the inflation is due to a combination of
population structure and ascertainment bias. The figure suggests that
the inflation of test statistics is approximately linear, and Devlin
and Roeder argued that this holds more generally. They therefore
proposed to calibrate the type I error of the Armitage test by
adjusting all test statistics by a constant factor $\lambda$. This
leaves the ranking of markers in terms of significance unchanged
[Figure~\ref{gcpc}(B)], and so GC is equivalent to adjusting the
significance threshold.

For most complex phenotypes, only a few genome-wide SNPs correspond to
strong causal associations, with test statistics in the upper tail of
the empirical distribution. Consequently, the bulk of the empirical
distribution, away from the upper tail, should reflect the null
distribution and can be used to estimate $\lambda$. Bacanu, Devlin and Roeder (\citeyear{PubMed10801388}) suggested estimating $\lambda$ as the ratio of the
empirical median to its null value ($=$0.455), because the median is
robust to a few large values in the upper tail. For the simulation of
Figure~\ref{gcpc}, the median of the test statistics is $0.59$, leading
to $\lambda=1.31$, a large value reflecting the strong ascertainment bias.

Setakis, Stirnadel and Balding (\citeyear{PubMed16354752}) pointed out that ascertainment bias can cause
median-adjusted GC to be very conservative. Marchini et~al. (\citeyear{PubMed15052271}) had
previously noticed that for strong population structure GC can be
anti-conservative when the number of test statistics used to estimate
$\lambda$ is $<$100, and conservative when the number is $\gg100$.
Devlin, Bacanu and Roeder (\citeyear{PubMed15514657}) ascribed this problem to failure to account for the
uncertainty in the estimate of $\lambda$, but Marchini et~al. (\citeyear{PubMed155146571a})
noted that this may not be the most important cause of the problem (see
also below). To allow for this uncertainty, Devlin, Bacanu and Roeder (\citeyear{PubMed15514657})
suggested using the mean of the test statistics to estimate $\lambda$,
since the mean-adjusted test statistics have an $F_{1,m}$ null
distribution. In the absence of true associations, Dadd, Weale and Lewis (\citeyear{PubMed19051284}) found mean-adjusted GC to be slightly superior to
median adjustment. However, the median is more robust to true positives
than the mean. As a compromise, Clayton et~al. (\citeyear{PubMed16228001}) proposed
adjusting on a trimmed mean, discarding say the highest $5\%$ or $10\%$
of test statistics.
\begin{lemma}
The mean of the smallest $100q\%$ values in a large random sample of
$\chi^2_1$ statistics has expected value
\[
\frac{1}{q}d_3(d_1^{-1}(q)),
\]
where $d_k$ is the distribution function of a $\chi^2_k$ random variable.
\end{lemma}
\begin{pf}
Let $X\sim\chi^2_1$, then
\begin{eqnarray*}
\mathbb{E} \bigl(X|X<d_1^{-1}(q) \bigr)
&=&\int_0^{d_1^{-1}(q)}x\frac{1}{q\sqrt{2\pi}}\frac{e^{-x/2}}{\sqrt{x}}\,dx\\
&=&\int_0^{d_1^{-1}(q)}\frac{1}{q\sqrt{2\pi}}{\sqrt{x}}e^{-x/2}\,dx\\
&=&\frac{1}{q}d_3(d_1^{-1}(q)).
\end{eqnarray*}
\upqed
\end{pf}

A limitation of all GC methods is that they do not distinguish markers
at which the pattern of association is correlated with the underlying
pedigree from those at which the pedigree does not contribute to the
association and so for which no adjustment should be necessary. Figure~\ref{gcpc}(B) and~(C) shows that median-GC-adjustment performs similarly to
PC-adjustment (see Section~\ref{pcr} below) in countering the overall
inflation of test statistics, but the corrected statistics can be very
different [Figure~\ref{gcpc}(D)] because PC-adjustment is SNP-specific.
GC often shows reduced power to detect association compared to rival
methods for adjusting for population structure.

In the remainder of this section we show connections between $\lambda$
and the kinship of study subjects. The Armitage test statistic can be
written as $T^2/V$, where $T$ is the difference between the allele
fractions in the samples of $n_1$ cases and $n_0$ controls,
\[
T=\sum_{i} \biggl(\frac{y_i}{n_1}-\frac{1-y_i}{n_0} \biggr)x_i,
\]
$V$ is an estimate of the variance of $T$,
\[
V= \biggl(\frac{1}{n_0}+\frac{1}{n_1} \biggr) \biggl(\frac{1}{n}\sum
_ix_i^2- \biggl[\frac{1}{n}\sum_ix_i \biggr]^2 \biggr),
\]
and $n=n_0+n_1$. In the following we assume retrospective
ascertainment, so that the case/control status $y$ is fixed by the
study design, while the allele count $x_i$ is random. Devlin and Roeder (\citeyear{PubMed11315092}) noticed that $\mathbb{E}[T]=0$, irrespective of
population structure, but that $\operatorname{Var}[T]$ can be inflated
relative to
$V$. In general,
%
%
\begin{eqnarray}\label{varform}
\hspace*{15pt}\operatorname{Var}[T]
&=&\sum_{i,j} \biggl(\frac{y_iy_j}{n_1^2}+\frac{(1-y_i)(1-y_j)}{n_0^2}\nonumber
\\[-8pt]\\[-8pt]
&&\hspace*{71pt}{}-\frac{(y_i-y_j)^2}{n_1n_0} \biggr)\nonumber
\\
&&{}\cdot\operatorname{Cov}(x_i,x_j),\nonumber
\end{eqnarray}
and substituting (\ref{cov}) into (\ref{varform}) leads to
\[
\operatorname{Var}[T]=\frac{4p(1-p)}{n_0n_1}(D+R),
\]
where
\begin{eqnarray}\label{Rpart}
\hspace*{20pt}D&=&\sum_{i} \biggl(\frac{n_0}{n_1}y_i+\frac{n_1}{n_0}(1-y_i)\biggr)K_{ii}\nonumber
\\
&\geq&\min\biggl(\frac{n_0}{n_1},\frac{n_1}{n_0} \biggr)\operatorname{Tr}[K],\nonumber\\
R&=&\sum_{i\neq j} \biggl(\frac{n_0}{n_1}y_iy_j+\frac{n_1}{n_0}(1-y_i)(1-y_j) \biggr)K_{ij}\nonumber
\\[-8pt]\\[-8pt]
&&{}-\sum_{i\neq j}(y_i-y_j)^2K_{ij}.\nonumber
\end{eqnarray}
It also follows that
%
%
\begin{equation}\label{ev}
\hspace*{10pt}\mathbb{E}[V]=\frac{4p(1-p)}{n_0n_1} \biggl(\sum_i K_{ii}-\frac{1}{n}\sum_{i,j}K_{ij} \biggr),
\end{equation}
which reduces to $2p(1-p)n/n_0n_1$ if all study subjects are outbred
and unrelated. Thus, provided that
%
%
\begin{equation}\label{limit}
\frac{V}{\mathbb{E}[V]} \stackrel{P}{\rightarrow} 1 \quad\mbox{as }
n\rightarrow\infty,
\end{equation}
we have
\[
\lambda=\mathbb{E} \biggl[\frac{T^2}{V} \biggr]\approx\frac{\operatorname{Var}
[T]}{\mathbb{E}[V]}=\frac{D+R}{\sum_iK_{ii}-(1/n)\sum_{i,j}K_{ij}}.
\]
Since $K$ is positive semi-definite, the second summation in (\ref{ev})
is $\ge0$, so that
\[
\lambda\geq\frac{D+R}{\operatorname{Tr}[K]}\geq\min\biggl(\frac{n_0}{n_1},\frac{n_1}{n_0} \biggr)+\frac{R}{\operatorname{Tr}[K]}.
\]
The dominant quantity bounding $\lambda$ is $R/\operatorname{Tr}[K]$
and since $\operatorname{Tr}
[K]\propto n$,
\[
\lambda\sim\frac{R}{\operatorname{Tr}[K]}\sim R/n.
\]
The large $n$ behavior of $\lambda$ depends on that of $R$. From~(\ref{Rpart}) we see that increasing levels of kinship either among cases or
among controls will tend to increase~$R$, while greater case-control
kinship tends to reduce~$R$. In the worst-case scenario, a typical
individual will be related at each degree of kinship to a fixed
proportion of the study sample as $n$ varies so that, unless the
average kinship among cases and among controls is balanced by the
average\break case-control kinship (such as when cases and controls are
matched within subpopulations under an island model), $R\propto n^2$ and
\[
\lambda\sim n,
\]
which generalizes the result of Devlin and Roeder (\citeyear{PubMed11315092}). In practice, it
is unclear how $\lambda$ should scale with $n$ in well designed GWAS
studies of homogeneous populations.

The statistic $T^2/\lambda V$ has the correct median or mean, but it
will not have an asymptotic $\chi^2_1$ null distribution unless (\ref{limit}) holds. This condition does not hold, for example, in an island
model with a fixed number of islands. This fact may underlie the poor
performance of GC in the settings discussed above. See Zheng et~al. (\citeyear{PubMed19432788}) for a detailed discussion of variance distortion in GC.

\subsection{Explicit Modeling of Genetic Correlations}\label{MCPsec}

GC is designed to correct the null distribution of a test statistic
derived under a probability model which is invalid in the presence of
structure. An alternative strategy is to derive a statistic from a
probability model that better reflects the actual data generating
process. For a retrospective study, the individual allele counts $x_i$
should be modeled as random variables satisfying (\ref{cov}). If we are
prepared to assume the higher order moments of $x$ are small, this can
be achieved with a regression model of the form (\ref{ret}) with linear
link and residuals
\[
x-\mathbf{1}\alpha-y\beta\sim N (0, \sigma^2K ).
\]
If we assign to $\alpha$ a diffuse Gaussian prior $N (0, \tau
^{-2} )$, with $\tau\downarrow0$, and to $\sigma^2$ the improper
(Jeffreys) prior with density proportional to $\sigma^{-2}$, we can
derive the score statistic $T^2/V$, with a $\chi^2_1$ asymptotic
distribution, where
\[
T=y^TPx
\]
and
\[
(n-1)V=y^TPy\cdot x^TPx-(y^TPx)^2
\]
for
%
%
\begin{equation}\label{pk}
P=K^{-1}-\frac{K^{-1}\mathbf{1}\mathbf{1}^TK^{-1}}{\mathbf{1}^TK^{-1}\mathbf{1}}.
\end{equation}

If the subjects are unrelated ($2K=I$), then $T$ reduces to a
comparison between the mean allele counts in cases and controls, as in
the Armitage test, but the variance $V$ is slightly smaller due to the
final term. When the relatedness between study subjects is unknown, as
in a typical GWAS, the estimate $\hat{K}$ of (\ref{corest}) may be
substituted for $K$. A similar approach, but with a different form for
$V$, has recently been proposed by Rakovski and Stram (\citeyear{RakovskiStram}), who point out
that when the kinships are known $T^2/V$ is equivalent to the QLS
statistic of Bourgain et~al. (\citeyear{PubMed12929084}).

The test described here could be used to analyze family data, either
using $K$ from the known pedigree or $\hat{K}$ estimated from genotype
data. For example, for pedigree-based $K$ in trios of two unrelated and
unaffected parents and an affected child, $T^2$ matches (up to a
constant) the numerator of (\ref{macstat}). $V$ differs slightly from
the denominator of (\ref{macstat}), reflecting the fact that the TDT
conditions on the parental genotypes, whereas the test described here
treats them as random. If the kinships are estimated from genome-wide
marker data, this test can exploit the between-family as well as the
within-family information, thus potentially increasing power over
FBTLA, while retaining protection from population structure. Moreover,
this is a very general approach, which applies for any ascertainment
scheme and degree of relatedness or population structure among study
subjects. Thus, if a researcher were unaware of the TDT, but applied
the retrospective regression model to family trio data, s/he would
automatically ``invent'' a test similar to the TDT but with potentially
superior properties.

\subsection{Structured Association}\label{struc}
Structured association (SA) methods are based on the island model of
population structure, and assume that the ancestry of each individual
is drawn from one or more of the ``islands.'' Popular software packages
include ADMIXMAP (Hoggart et~al., \citeyear{PubMed12817591}) and STRUCTURE/STRAT (Pritchard and Donnelly, \citeyear{PubMed11855957};
Falush, Stephens and Pritchard,\break \citeyear{PubMed12930761}). These approaches model variation in
ancestral subpopulation along a chromosome as a Markov process.
Stratified tests for association (Clayton,\break \citeyear{clayhbook}), such as the
Mantel--Haenszel test, can then be performed to combine signals of
association across subpopulations. More generally, a logistic
regression model of the form (\ref{pro}) can be employed, with
admixture proportions (one for each subpopulation) entering as covariates.

Similar to GC, SA methods can be effective using only $\sim$10$^2$ SNPs,
but unlike GC, they can be computationally intensive, although a
simplified and fast version of SA is implemented in PLINK (Purcell et~al., \citeyear{PubMed17701901}). The number of subpopulations can be estimated from
the data by optimizing a measure of model goodness of fit, but this
increases the computational burden and there is usually no satisfactory
estimate because, as we noted above in Section~\ref{popstruc}, the
island model is not well suited to most human populations. Indeed,
ADMIXMAP was primarily designed for admixture mapping, in which the
genomes of admixed individuals are scanned for loci at which cases show
an excess of ancestry from one of the founder populations (McKeigue, \citeyear{McKHBook}). Because of the limited number of generations since the
admixture event, this approach has features in common with linkage as
well as association study designs.

\subsection{Regression Control}
Wang, Localio and Rebbeck (\citeyear{PubMed15920341}) showed that it is possible to control for
population structure within a logistic regression model of the form
(\ref{pro}) by including among the covariates the genotype at a single
marker that is informative about ancestry. Setakis, Stirnadel and Balding (\citeyear{PubMed16354752})
proposed using a set of $\sim$10$^2$ widely-spaced SNPs, which are
assumed to be noncausal (in practice, a ``random'' set of SNPs). These
null SNPs are informative about the underlying pedigree, which we have
argued forms the basis of the problem of inflation of test statistics
due to population structure. Including these SNPs as regression
covariates while testing a SNP of interest should eliminate most or all
of the pedigree (population structure) effect.

Setakis, Stirnadel and Balding (\citeyear{PubMed16354752}) suggest two standard procedures to avoid
overfitting the SNP covariates: a backward (stepwise) selection and a
shrinkage penalty approach. In the absence of ascertainment bias, both
methods performed similarly to GC and SA, while being computationally
fast and allowing the flexibility of the regression framework. With
ascertainment bias, the regression control approach substantially
outperformed GC.

Another approach (Epstein, Allen and Satten, \citeyear{PubMed17436246}), which is also related to
propensity score methods, uses ancestry-informative SNPs to create a
risk score, stratifies study subjects according to this score, and
performs a stratified test of association.

\subsection{Principal Component Adjustment}\label{pcr}
Zhang, Zhu and Zhao (\citeyear{PubMed12508255}) proposed controlling for population structure in
quantitative trait association analysis by including principal
components (PCs) of genome-wide SNP genotypes as regression covariates.
Price et~al. (\citeyear{PubMed16862161}) presented a similar method, focusing on its
application to case-control GWAS. PC regression is similar to the
regression control approach of Setakis, Stirnadel and Balding (\citeyear{PubMed16354752}), but minimizes
overfitting by using only a few linear combinations of SNPs (the PCs),
rather than a larger number of individual SNPs. However, many more SNPs
(typically $\sim$10$^4$) are required in the PC-based approach.

Let $X$ denote a matrix with $n$ rows corresponding to individuals and
$L$ columns corresponding to SNPs. Genotypes are initially coded as
allele counts (0, 1 or 2) but are then standardized to have zero mean
and unit variance. Then the $n\times n$ matrix $XX^T/L$ is the
estimated kinship matrix $\hat{K}$ introduced at (\ref{corest}). Since
$\hat{K}$ is symmetric and positive semi-definite, it has an eigenvalue
decomposition
%
%
\begin{equation}\label{eigvaldecomp}
\hat{K}=\frac{1}{L}XX^T=v\Lambda v^T,
\end{equation}
where the columns of $v$ are the eigenvectors, or PCs, of $\hat{K}$,
while $\Lambda$ is a diagonal matrix of corresponding (nonnegative)
eigenvalues in decreasing order.

Standard principal components analysis uses the $L\times L$ matrix
$X^TX$ specifying the empirical correlations between the columns of the
design matrix. Here, the variables of interest are the individuals,
corresponding to rows, and, hence, we focus on $\hat{K}=XX^T/L$.
However, because $X$ is column-standardized, and not row-standardized,
$\hat{K}$ is not an empirical correlation matrix. In particular, the
diagonal entries of $X^TX$ are all one, whereas the diagonal entries of
$\hat{K}$ vary over individuals according to their estimated inbreeding
coefficient (Section~\ref{secrelation}).

To maximize the empirical variance of $X^Tv_1$, the first PC $v_1$ will
typically be correlated with many SNPs. For example, in a two-island
model, it will have greatest correlation with the SNPs whose allele
fractions are most discrepant between the two islands. Thus, $v_1$ acts
as a strong predictor of island membership, and can also identify
admixed individuals (intermediate scores). More generally, in an
$S$-island model the first $S-1$ PCs predict island memberships and
individual admixture proportions (Patterson, Price and Reich, \citeyear{PubMed17194218}) (see Figure
\ref{pcplot} for an illustration with $S=3$). The subsequent PCs
represent the within-island pedigree effects. Cryptic kinship typically
generates weaker LD than large-scale population structure, and the
effects of small groups of close (even first-degree) relatives are
usually not reflected in the leading PCs.

Tightly linked SNPs tend to be in high LD with each other, and
sometimes one or more of the leading PCs will be dominated by large LD
blocks. Since such blocks are genomically local, they convey little if
any information about population structure. One way to avoid this
problem is to filter GWAS SNPs prior to extracting the PCs, to exclude
one in each pair of high LD SNPs.

%
\begin{figure}

\includegraphics{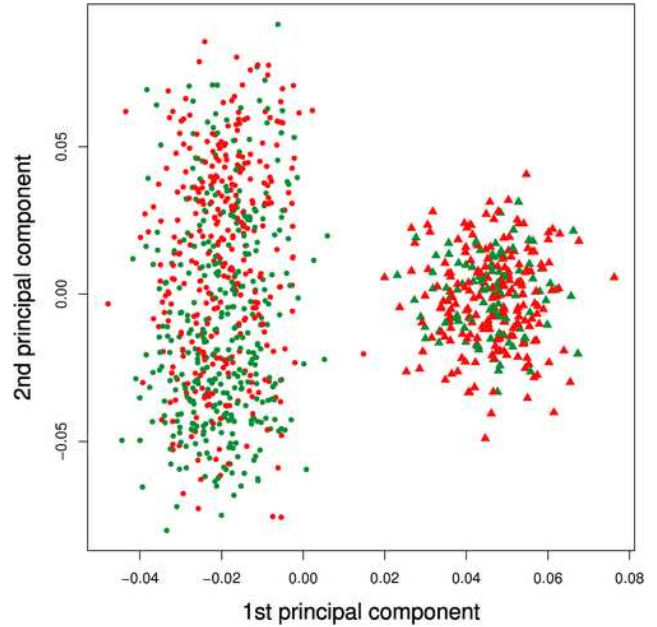}

\caption{First two principal component scores for the
cases (green) and controls (red) of the simulation underlying Figure~\protect\ref{gcpc}.
Triangles indicate individuals from Island 1, and circles
the admixed individuals with ancestry from Islands 2 and 3.}\label{pcplot}
\end{figure}

As for structured association and regression control, the idea
motivating PC adjustment is that if a correlation between the phenotype
and the tested SNP can be partly explained by measures of ancestry,
here PCs, then including these as regression covariates prevents that
part of the signal from contributing to the test statistic. In
particular, in the case of linear regression, only the components of
the phenotype and genotype vectors that are orthogonal to the PCs
included in the model contribute to the test statistic. This should
protect against spurious associations, provided that sufficient PCs are
included in the model to explain potentially confounding structure. For
example, in the population of Figure~\ref{pcplot}, $v_1$ and $v_2$ can,
with accuracy, jointly predict the proportion of an individual's
ancestry arising from each of the three subpopulations, and because of
the varying case-control ratios across the subpopulations, they can
also to some extent predict case-control status. Including the PCs as
covariates in the regression model discards information, for example,
indicating that alleles common in the high-risk subpopulation are more
likely to be causal. Because of the danger of ascertainment bias in
retrospective studies, such information may be dangerous and it is
safer to discard it, but if ascertainment is not a problem, for
example, in a prospective study, discarding this information is inefficient.

For computational reasons the EIGENSTRAT software (Price et~al., \citeyear{PubMed16862161}), which implements PC adjustment, does not include PCs
as logistic regression covariates, but instead uses a linear adjustment
of both phenotypes and genotypes. Such an adjustment is valid only
under the assumption that the $y_i$ form a homoskedastic sample (Agresti, \citeyear{Agresti}, page~120) and should be reasonable if the sample case:control
ratio is not too far from 1, and the effect sizes are small.

By default, the EIGENSTRAT software includes the first ten PCs. Patterson, Price and Reich (\citeyear{PubMed17194218}) proposed a test to determine whether the lead
eigenvalue of $\hat{K}$ is significantly larger than one would expect
under a null model, but it remains unclear what significance threshold
for this test might be appropriate, if any, for protecting association
test statistics from inflation. As for any other regression covariate,
there is an argument for only including a PC in the model if it shows
an association with the phenotype (Novembre and Stephens, \citeyear{PubMed18425127};
Lee, Wright and Zou, \citeyear{zoupreprint}).
Experience seems to suggest that between 2 and 15 PCs are typically
sufficient, and in large studies for which $n$ may be several thousand,
these will correspond to a small loss of total genotypic information.

While the intuition motivating PC-adjustment is valid under an island
model, protection from population structure effects is not guaranteed
under more complicated and realistic models of population structure. In
particular, inflation of test statistics due to cryptic relatedness is
not ameliorated by PC adjustment. Moreover, if leading PCs reflect
genome-local effects, PC adjustment could lose valuable information and
lead to true effects being missed.

\subsection{Mixed Regression Models}

We assume here a quantitative phenotype with $g$ the identity link. The
linear mixed model (MM) extends (\ref{pro}) by including for each
individual $i$ a latent variable $\delta_i$ such that
%
%
\begin{equation}\label{mixed2}
\mathbb{E}[y_i|\delta_i]=\mathbb\alpha+x_i\beta+\delta_i.
\end{equation}
The value of $\delta_i$ is interpreted as a polygenic contribution to
the phenotype, due to many small, additive, genetic effects distributed
across the genome. In animal breeding genetics, the equivalent term is
referred to as the \textit{breeding value}. The additive assumption seems
to be well supported for traits with a complex genetic basis (Hill, Goddard and Visscher, \citeyear{PubMed18454194}),
although it is also possible to include latent
variables corresponding to dominance effects. Under the additive
polygenic assumption, the variance-covariance structure of $\delta$ is
proportional to the correlation structure of genotypes coded as allele
counts, which from (\ref{cov}) is proportional to $K$, the kinship
matrix. Hence, we assume
%
%
\begin{equation}\label{del}
\delta\sim N (0,2\sigma^2h^2K ),
\end{equation}
where $h^2\in[0,1]$ is the \textit{narrow sense heritability}, and is
defined as the proportion of phenotypic variation that can be
attributed to additive polygenic effects. The residuals are assumed to satisfy
\[
y_i-\mathbb\alpha-x_i\beta-\delta_i \sim N \bigl(0,\sigma^2(1-h^2)I \bigr).
\]
The origins of this linear MM lie in the partitioning by Fisher (\citeyear{fisher1918}) of the variance of a quantitative trait into independent
genetic and environmental components, and derivation of the genetic
correlation of trait values of a pair of relatives assuming Mendelian
inheritance (Gianola, \citeyear{hbmixed}). It is conventional to introduce the 2 in
(\ref{del}) because $2K$ reduces to $I$ in the limiting case of
completely unrelated and completely outbred individuals, in which case
$h^2$ becomes inestimable.

The model (\ref{mixed2}) has long been used for mapping quantitative
trait loci in outbred pedigrees, using a pedigree-based kinship matrix
(H{\"{o}}schele, \citeyear{hbQTL}). Yu et~al. (\citeyear{PubMed16380716}), who were interested in
association mapping in maize, were first to suggest using the same
model to correct association analysis for population structure, but
with $K$ estimated from marker data. In fact, Yu et al. also include
in their model additional population structure terms, namely, the
ancestral proportions estimated by the STRUCTURE software (Section~\ref{struc}).
Zhao et~al. (\citeyear{PubMed17238287}) used the same model to analyze an
\textit{Arabidopsis thaliana} data set and, like Yu et al., found that
the additional terms improved the fit of the model. However, neither
set of authors formally assess the improvements in fit which, by visual
inspection of the genome-wide $p$-value distributions, seem modest. In
principle, $K$ already includes population structure information,
making the additional terms redundant. However, the structure terms
provide a low-dimensional summary of key features of $K$ which are
likely to be better estimated than individual kinships, and this may
generate some advantage to including the structure terms in the
regression model as well as $K$. Typically, $\sim$10$^5$ SNPs are
required for adequate estimation of $K$ in human populations, more than
are required for estimation of PCs, but this number is usually
available from a GWAS.

Kang et~al. (\citeyear{PubMed18385116}) have developed the software package EMMA for
fast inference in linear MMs using a likelihood ratio test.
Alternatively, for very large samples one can use the score test which
is computationally faster because it only requires parameter estimates
under the null hypothesis ($\beta=0$). Another fast method for
inference in mixed models, GRAMMAR, has been proposed by Aulchenko, de Koning and Haley (\citeyear{PubMed17660554}). Although GRAMMAR is faster than EMMA, it is an
approximate method and the authors found that it could be conservative
and hence have reduced power. GRAMMAR uses the mixed model to predict
the phenotype under the null hypothesis,
\[
\hat{y}=\hat{\alpha}+\hat{\delta},
\]
where $\hat{\delta}$ is the best linear unbiased predictor (BLUP) of
$\delta$, which is equivalent to the empirical Bayes estimate for
$\delta$ with prior (\ref{del}) (Robinson, \citeyear{RobinsonBLUP}). This prediction
only needs to be made once for the whole data set. The next step is to
use the residuals from the prediction as the outcome in a linear regression,
%
%
\begin{equation}\label{GRAMMARreg}
y-\hat{y}=\mathbf{1}\mu+x\beta+\varepsilon,
\end{equation}
and test the parameter $\beta$ for each SNP. An assumption underlying
(\ref{GRAMMARreg}) is that the residuals are independent and
identically distributed, which is strictly false. Indeed, both $\mathbb{E}(y-\hat{y})$ and $\operatorname{Var}(y-\hat{y})$ are
functions of $K$ unless
$h^2=0$, which may be the reason that an additional GC-style variance
inflation correction is often required to calibrate the GRAMMAR type I
error rate.

Note that (\ref{mixed2}) can be reparametrized as
\[
\mathbb{E}[y_i|\gamma]=\mathbb\alpha+x_i\beta+v_i\gamma,
\]
with
\[
\gamma\sim N (0,2\sigma^2h^2\Lambda),
\]
where $\Lambda$ and $v_i$ are defined above at (\ref{eigvaldecomp}).
Thus, the MM approach uses the same latent variables as PC adjustment
but deals differently with the vector $\gamma$ of nuisance parameters.
From a Bayesian point of view, both methods put independent priors on
the components of $\gamma$. Using $k$ PCs as regression covariates can
be viewed as assigning to each of the $n-k$ trailing components of
$\gamma$ a prior with unit mass at zero, while each of the $k$ leading
components receives a diffuse prior. These assignments imply certainty
that the polygenic component of the phenotype is fully captured by the
first $k$ PCs. In contrast, the MM approach puts a Gaussian prior on
each component of $\gamma$, with variances proportional to the
corresponding eigenvalues.

\section{Simulations}\label{sim}
\subsection{Case-Control Studies Without\break Ascertainment Bias}
We show here the results of a small simulation study designed to
illustrate the merits of some of the methods introduced above for
correcting GWAS analysis for population structure. Although we have
criticized the island model as unrealistic, it remains the most
convenient starting point and we use it below, the only complication
considered here being ascertainment bias. More extensive and realistic
simulations will be published elsewhere.

We simulated data from $500$ case-control studies, each with 1000
cases and 1000 controls drawn from a population of 6000
individuals partitioned into three equal-size subpopulations. Ancestral
minor allele fractions were $\operatorname{Uniform}[0.05,0.5]$ for all $10{,}000$
unlinked SNPs. For each SNP, we drew subpopulation allele fractions
from the beta-binomial model described in Balding and Nichols (\citeyear{PubMed7607457}). Under
this model, a marker with ancestral population allele fraction $p$ has
subpopulation allele fractions that are independent draws from
\[
\operatorname{Beta} \biggl(\frac{1-F}{F}p,\frac{1-F}{F}(1-p) \biggr),
\]
where $F$ is Wright's $F_{ST}$, a measure of population divergence
(Balding, \citeyear{PubMed12689793}). In order to discriminate among the methods, we
simulated a high level of population structure, $F=0.1$, which is close
to\break between-continent levels of human differentiation; this is larger
than is typical for a well-designed GWAS, but some meta-analyses may
include populations at this level of differentiation. The studies
simulated here are relatively small by the standards of current GWAS,
and larger studies will be affected by less pronounced structure than
that simulated here.

We simulated the disease phenotype under a logistic regression model,
with $20$ SNP markers each assigned allelic odds ratio $1.18$. The
population disease prevalence was $0.18$. We performed tests for
disease-phenotype association for all the markers in each data set,
using median-adjusted GC, principal component adjustment with 10 PCs
(PC10) and the likelihood ratio test from the linear mixed model (MM).
Note that the PC10 and MM approaches apply a linear regression model to
binary outcome data, which (as noted in Section~\ref{pcr}) should be
reasonable if the case:control ratio is not extreme and the effect
sizes are small. We also performed the score test described in Section~\ref{MCPsec}. This retrospective model is consistent with the
case-control ascertainment, although in fact the resulting score
statistic is symmetric in $x$ and $y$. We call this test MCP to stand
for Multivariate Gaussian model Conditional on
Phenotype.

%
\begin{figure}

\includegraphics{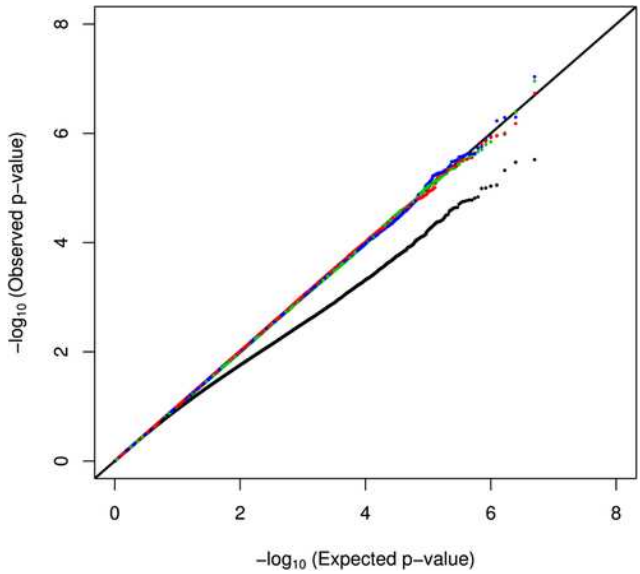}

\caption{Expected versus observed
$-\log10$ ($p$-values) for four test statics, GC (black), PC10 (blue), MM
(red) and MCP (green) evaluated at 5 million SNPs (10,000 per dataset)
simulated under the null of no association in population case-control
studies (see text for details of the simulation).}\label{logQQbinret}
\end{figure}

The PC10, MM and MCP approaches all require specification of $K$, the
kinship matrix of the study subjects. Figure~\ref{logQQbinret} shows
the observed and expected $p$-values at the null markers, aggregated
over the $500$ simulations, using $\hat{K}$ estimated via genetic
correlations (\ref{corest}). We see that the type I error is well
calibrated for all the methods except GC, which has a conservative null
distribution. We repeated the analysis using the true $K$ used for the
simulation and the results are similar (not shown). GC is conservative
because in this extreme scenario with large $F_{ST}$ and a small number
of subpopulations, the assumption that the test statistic
asymptotically follows a linear-inflated $\chi^2_1$ distribution fails.
The condition (\ref{limit}) is not satisfied in the beta-binomial model
with fixed $F_{ST}$ unless the number of subpopulations increases with $n$.

%
\begin{figure*}

\includegraphics{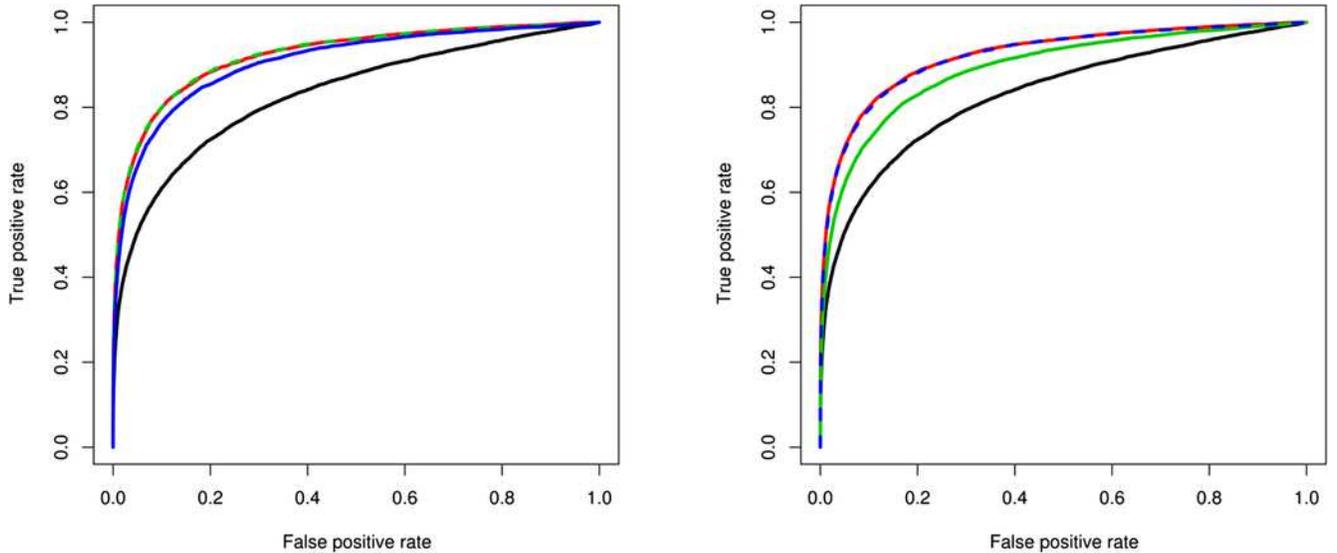}

\caption{ROC plots comparing true and false-positive
rates of GC (black), PC10 (blue), MM (red) and MCP (green) estimated
from data aggregated over 500 simulated retrospective population
case-control studies. Left: true kinships, $K$; right: estimated
kinships, $\hat{K}$.}\label{binretROC}
\end{figure*}

In order to compare power across the methods, we plotted ROC curves for
the statistics from the four methods using both the true $K$ and the
estimated $\hat{K}$ (Figure~\ref{binretROC}). GC has lower power than
the other three methods, even though the ROC calibrates its bad false
positive rate. When $K$ is used, the MCP and MM approaches are equally
powerful, and more powerful than the other two methods, because both
exploit the between-subpopulation information. When $\hat{K}$ is used
both the MM and MCP methods lose their power advantage over PC10, which
may be due to the sampling error in the eigenvectors of $\hat{K}$. PC
correction uses only the leading eigenvectors to adjust the analysis;
these are less affected by noise and for an island model they contain
all the population structure signal. For an actual GWAS, the MM and MCP
methods should show performance somewhere intermediate between the
cases considered here, because many more than 10,000 SNPs are available
to estimate $K$.

\subsection{Case-Control Studies With Ascertainment Bias}
We repeated the simulation studies described above but with
ascertainment bias. Specifically, each simulated study sampled $50$
controls from two of the three subpopulations and $900$ controls from
the remaining subpopulation. This corresponds to a scenario in which
investigators are forced to search\break widely across subpopulations to
obtain a sufficient number of cases for their study, but are able to
recruit most controls from the local subpopulation. The case:control
ratio varies dramatically over subpopulations in this scenario, so that
subpopulation allele frequencies are strong predictors of case-control status.

Once again, PC10, MM and MCP all have good control of type I error,
while GC is dramatically conservative (not shown). We again plotted ROC
curves for the statistics from the four methods using both $K$ and
$\hat{K}$ (Figure~\ref{binretascROC}). GC shows almost no power in these
simulations. When $K$ is used the MCP and MM approaches are equally
powerful and more powerful than PC10. Their power advantage over PC10
is more substantial than in the previous scenario without ascertainment
bias, because here the leading eigenvectors of $K$ are stronger
predictors of case-control status. When $\hat{K}$ is used the MCP and
MM methods again lose some power compared with PC10 and again the MCP
method suffers a greater loss than the MM test.

%
\begin{figure*}

\includegraphics{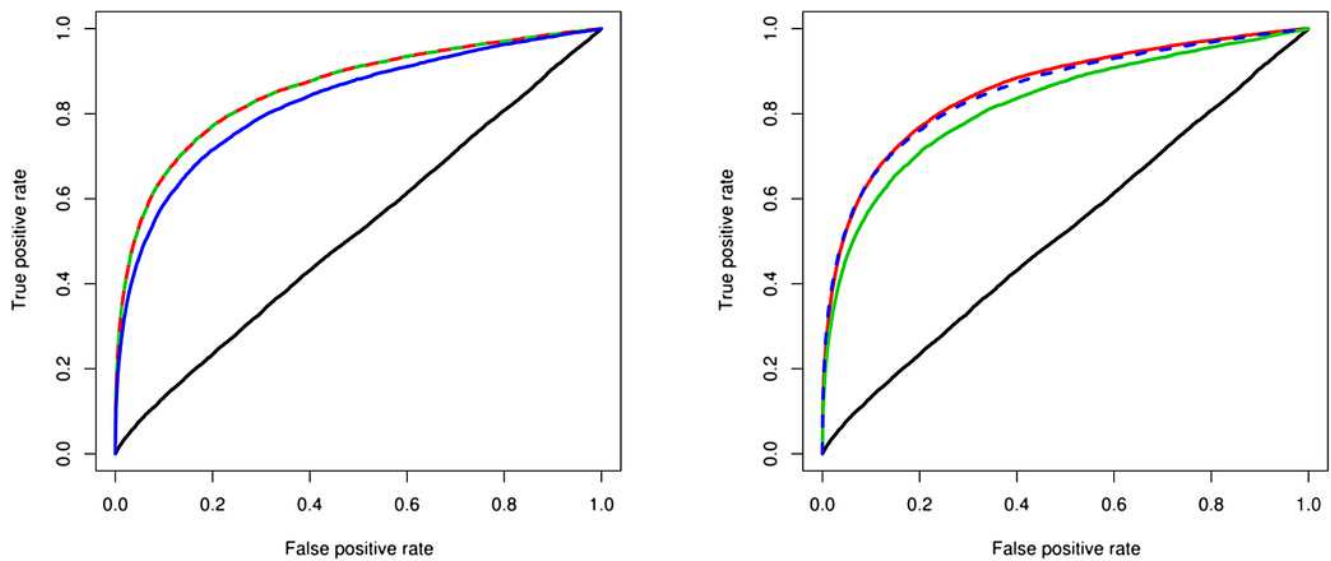}

\caption{ROC plots comparing true and
false-positive rates of GC (black), PC10 (blue), MM (red) and MCP
(green) estimated from data aggregated over 500 simulated population
case-control studies with biased ascertainment of controls. Left: true
kinships, $K$; right: estimated kinships, $\hat{K}$.}\label{binretascROC}
\end{figure*}

\section{Summary}
The theme of our review has been the unifying role of the matrix of
kinship coefficients, $K$, and its estimate $\hat{K}$ defined at (\ref{corest}). We view population structure and cryptic kinship as the
extremes of the same confounder, the latent pedigree, and $\hat{K}$ as
a good summary of the pedigree for use in adjusting association
analyses. We have also argued that, whereas methods are often tested
under an island model of population structure, these models do not
provide realistic descriptions of relatedness in human populations.

The median-adjusted Genomic Control (GC) is simple to apply and, for
association studies with moderate sample sizes and small amounts of
within sample relatedness, it is a satisfactory method for protecting
against confounding, which requires relatively few SNPs ($\sim$10$^2$).
When the study sample is drawn from a population with a few, distinct
subpopulations, GC should be used with caution because the $\chi^2$
approximation may fail. Our simulations also confirmed previous reports
that GC is very sensitive to ascertainment bias. Structured association
and regression adjustment may also be used with relatively few SNPs,
and the latter has important advantages over GC.

Linear mixed models (MM) and the multivariate Gaussian model
conditioning on phenotype (MCP) explicitly model genetic correlations
using $K$, and are respectively appropriate for prospective and
retrospective studies with genome-wide SNP data. They are particularly
suited to modeling complex patterns of kinship, including intermediate
scenarios between close relationships and large scale structure, which
can arise in plant genetics, human population isolates and in animal
breeding studies. Previous computational limitations have largely been
overcome in recent years. These models also provide powerful and
computationally efficient methods for analyzing family data of any
structure, and combinations of families and apparently unrelated individuals.

Principal components (PC) adjustment in effect eliminates from test
statistics the part of the phenotypic signal that can be predicted from
large scale population structure. In particular, if natural selection
leads to large allele frequency differences across subpopulations at a
particular SNP, and case-control ratios vary across subpopulations,
then spurious associations can arise that PC adjustment will control,
because the SNP genotypes are strongly correlated with subpopulation,
whereas the MM and MCP methods will not. On the other hand,
the MM and MCP methods can gain power
over PC adjustment because they explicitly model the phenotype-genotype
correlations induced by relatedness and genetic drift. For example,
they should provide better power than PC adjustment when analyzing data
from human population isolates which are homogeneous for environmental
exposures.

There are probably better ways to extract the signals of population
structure from an estimate of $K$ than those considered here. Selection
of the first few principal components of $\hat{K}$ can be viewed as a
form of signal denoising, but PC regression adjustment does not in
general optimize power. It may be possible to adapt the MM approach to
maintain the power advantage over PC while reducing noise, for example,
by smoothing or truncating the lower order eigenvalues of $\hat{K}$.

\section*{Acknowledgments}
We gratefully acknowledge helpful
discussions from many colleagues, including Yurii Aulchenko, David
Clayton, Clive Hoggart, Chris Holmes, Matti Pirinen, Sylvia Richardson
and Jon White, and, in particular, Francois Balloux, Dan Stram and Mike
Weale for helpful comments on an early draft.
This work was supported in part by the UK Medical Research Council and GlaxoSmithKline.

\vspace*{-2.2pt}
\end{document}